\documentclass[%
 aip,
 apl,
 amsmath,amssymb,
 reprint,%
]{revtex4-1}

\usepackage{graphicx}% Include figure files
\usepackage{dcolumn}% Align table columns on decimal point
\usepackage{bm}% bold math
\usepackage{physics}% Physics by SJF
\usepackage[utf8]{inputenc}
\usepackage[T1]{fontenc}
\usepackage{mathptmx}
\usepackage{etoolbox}

\makeatletter
\def\@email#1#2{%
 \endgroup
 \patchcmd{\titleblock@produce}
  {\frontmatter@RRAPformat}
  {\frontmatter@RRAPformat{\produce@RRAP{*#1\href{mailto:#2}{#2}}}\frontmatter@RRAPformat}
  {}{}
}%
\makeatother
\begin{document}

\preprint{AIP/123-QED}

\title{High mobility holes at germanane/Ge(111) allotropic cross-dimensional heterointerface}
% Force line breaks with \\
\author{Yumiko Katayama}
%\email{katayama@g.ecc.u-tokyo.ac.jp}
\affiliation{Graduate School of Arts and Sciences, University of Tokyo, Komaba, Meguro, Tokyo 153-8902, Japan}
\author{Daiki Kobayashi}
\affiliation{Graduate School of Arts and Sciences, University of Tokyo, Komaba, Meguro, Tokyo 153-8902, Japan}
\author{Hikaru Okuma}
\affiliation{Graduate School of Arts and Sciences, University of Tokyo, Komaba, Meguro, Tokyo 153-8902, Japan}
\author{Yuhsuke Yasutake}
\affiliation{Graduate School of Arts and Sciences, University of Tokyo, Komaba, Meguro, Tokyo 153-8902, Japan}
\author{Susumu Fukatsu}
\affiliation{Graduate School of Arts and Sciences, University of Tokyo, Komaba, Meguro, Tokyo 153-8902, Japan}
\author{Kazunori Ueno}
\email{ueno@phys.c.u-tokyo.ac.jp}
\affiliation{Graduate School of Arts and Sciences, University of Tokyo, Komaba, Meguro, Tokyo 153-8902, Japan}

\date{\today}% It is always \today, today,

\begin{abstract}
Germanane (GeH) is essentially a hydrogen-terminated Ge analog of graphene with a direct gap ($\approx$1.6 eV). 
Record hole mobility $\mu_{\rm h}\!\approx\,$67,000 cm$^{2}$V$^{-1}$s$^{-1}$ is found at 15 K for a single allotropic cross-dimensional(D) heterointerface.
This is enabled by making topotactically-transformed 2D GeH layers meet the 3D bulk Ge(111). 
Temperature dependence of $\mu_{\rm h}$ implies metallic conduction without ionized impurity scattering between 20 K and 250 K.
Sheet hole density for a Fermi sphere $n_{\rm S}\!=\!2.8\times10^{11}$ cm$^{-2}$ agrees well with $3.0\times10^{11} {\rm cm}^{-2}$ of Hall measurements. 
A 6,500-\% magnetoresistance at 7 T accompanies Shubnikov-de Haas oscillations visible even at 15 K. 
These imply single-band conduction of holes with small effective mass in the in-plane directions, invoking
a 2D hole gas (2DHG) picture that allotropic cross-D heterointerface between 2D GeH and 3D Ge harbors 2D-confined high-mobility holes.
Even without elaborate heteroepitaxy and modulation doping, allotropic cross-D heterostructures pave the way toward  facile 2DHG creation.
\end{abstract}

\maketitle

%\section{\label{sec:level1}First-level heading:\protect\\ The line
%break was forced \lowercase{via} \textbackslash\textbackslash}

Advanced electronics and the evolving research fields in solid-state physics arguably build on high-mobility materials. 
They include man-made compounds like GaAs and two dimensional(2D) van der Waals semiconductors\cite{1,2,3,4,5,6,7,8}. 
The semi-naturally-occurring graphene touts an exceptionally high mobility arising from its gapless, Dirac-point dispersion. 
Allied 2D crystalline materials have garnered renewed interest from the viewpoint not only from physics but high-speed electronics in the hope to find a gapped one that allows for efficient switching.

High mobilities, elusive in the bulk due to ionized impurity scattering, are achievable with 2D electron(hole) gas (2DE(H)G) localized by design at heterointerfaces like Al$_x$Ga$_{1-x}$As/GaAs\cite{9,10,11,12,13,14,15}. 
Meanwhile, albeit gapless, graphene's exfoliability and stackability have technological impacts on electronics as they could bring a gap-engineered heterostructure when combined with 2D systems like transition metal dichalcogenidesBO and black phosphorus (BP). 
Topological semimetals  \cite{16,17} hold promise for high mobilities with added advantages of valleytronics\cite{18} and spintronics\cite{19}. 
Recently, the focus has shifted to stacked interfaces like narrow-gap BP and hexagonal boron nitride (h-BN) \cite{15,20,21,22}.
This aligns with the fact that high-mobility \textit{heterointerfaces} in the literature involve a multipartite, largely bipartite, entry of 3D materials. 
However, such a notion is defied when "2D meets 3D" as the prefix "hetero" is overridden to enhance mobility. 
Here we demonstrate that allotropic, in between \emph{homo} and  \emph{hetero}, cross-D heterointerfaces are facile yet potentially useful and even game-changing.

Germanane (GeH) is a 2D semiconductor, essentially a stacked array of van der Waals sheets due to vertically hydrogen-terminated honeycomb germanium lattice \cite{23}. 
Unlike graphene and group-IV analogs such as silicene and germanene GeH has a direct gap $E_{\rm g}\!\approx$1.6 eV \cite{24,25} and a smaller electron effective mass than that of Ge. 
GeH can be synthesized from crystalline CaGe$_2$ by topochemical intercalation, with Ca replaced by H in wet processes using HCl or HF \cite{23,24,26}. 
GeH and molecular-engineered germanane modified with functional groups \cite{27,28,29,30,31} offer a vast range of applications including photocatalysis \cite{32}, photoelectrochemical photodetector \cite{33}, anode for lithium-ion batteries \cite{34}, and field-effect transistors (FETs) \cite{35,36,37,38}. 
Previously, the authors reported the ambipolar action of a GeH thin-film FET on Ge(111) substrate in an electric double-layer transistor (EDLT) configuration. Hall mobilities ($\mu_{\rm H}$) of electrons and holes at 120 K were considerably high, 6500 and 570 cm$^2$V$^{-1}$s$^{-1}$, respectively\cite{36}. 
However, the transport properties at cryogenic temperatures remained unexplored due to soaring contact resistances with lowering $T$. 
As such little has been clarified of the high mobility channels. 
In the anticipation that only GeH is relevant, even the first "2D meeting 3D" might have been missed.

Here an attempt is made to explore conductive channels at the allotropic cross-D heterointerface (AXI) between GeH and Ge based on low-$T$ measurements.
Larger-than-expected hole mobilities up to 67,000 cm$^2$V$^{-1}$s$^{-1}$ and clear Shubnikov-de Haas (SdH) oscillations are observed. 
Comparable sheet carrier densities after SdH oscillations and Hall effects imply single-band hole conduction. 
Additionally, $T$-dependence of $\mu_{\rm H}$ suggests a suppressed ionized acceptor scattering.
These imply the relevance of 2DHG that spontaneously forms near the GeH/Ge(111) AXI without modulation doping.

%\vspace{-2mm}
%\section*{\large{Results and Discussion}}

\begin{figure}[!b]
\begin{center}
\vspace{-3mm}
\includegraphics[width =0.83 \linewidth]{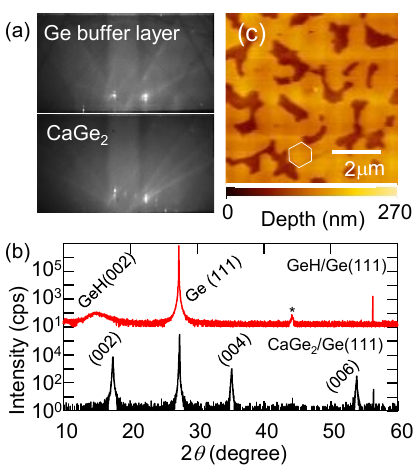}
\end{center}
\vspace{-4mm}
\caption{(a) RHEED patterns of Ge buffer and CaGe$_2$ layer. (b) XRD 2$\theta/\omega$ scans of the CaGe$_2$ film and the GeH thin film on Ge (111) substrate. 
The weak peak at 2$\theta$= 56.29 is due to the Ge(222) diffraction forbidden by the extinction rule.
The spurious feature around 44 degrees (marked by asterisk) is noted, which is due to the stainless-steel plate of the equipment, i.e., Fe(110). (c) AFM image of the GeH film.  }
\vspace{-4mm}
\label{fig1}
\end{figure}

Crystallinity and surface morphology of GeH/Ge(111) thin films are discussed first. 
Epitaxial CaGe$_2$ were initially grown on the Ge buffer over a Ge(111) substrate followed by topochemical transformation \cite{36,38}. 
This was confirmed \textit{in situ} by reflection high energy electron diffraction (RHEED) (Fig.\,\ref{fig1}(a)). 
X-ray 2$\theta/\omega$ scans on the CaGe$_2$ films had sharp peaks (Fig.\,\ref{fig1}(b)). 
The FWHM of the rocking curve of the (002) peak measures $\approx$ 0.08 degrees. 
Topochemical reaction brought a broad one at 2$\theta\!=\!14.9$ degrees on the upper trace.
This is the $(002)$-diffraction due to GeH. 
The broadening is presumably caused by fluctuations in the GeH interlayer distance during topochemical reaction\cite{23,24}. 
Assuming the 2H structure, the lattice constant $c\!=\!11.9$\,\r{A} is found. 
The monolayer GeH thickness is estimated to be 5.9\,\r{A}, in good agreement with the literature\cite{23,24,36,38,39}. 
The atomic force microscopy (AFM) in Fig.\,\ref{fig1}(c) reveals the crystalline GeH film of micrograins $\approx1\,\mu$m.
Their six-fold symmetry about [00l] is recognized with reference to the white hexagon as an eye-guide. 
Coalescence of hexagonal prismatic crystals could contribute to the GeH/Ge(111) conducting channel discussed later.
Meanwhile, a high degree of crystallinity of CaGe$_2$ implies as good quality across the GeH layer.

\begin{figure}[!b]
\vspace{-3mm}
\begin{center}
\includegraphics[width =0.8 \linewidth]{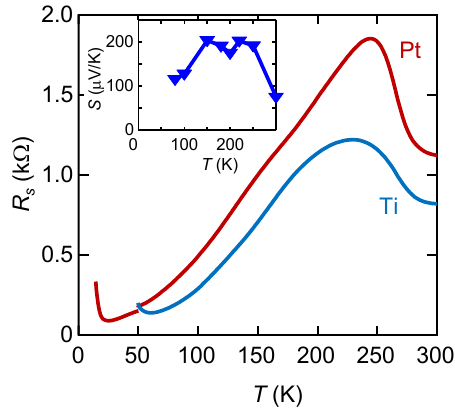}
\end{center}
\vspace{-3mm}
\caption{Temperature dependence of sheet resistance $R_{\rm s}$ of GeH/Ge(111) thin films with Pt (red line) and Ti (blue line) electrodes attached. 
Four-terminal resistances are plotted in the temperature range where two-terminal resistances are smaller than 100 k$\Omega$.
For this reason, the accessible temperature is lower, $15$\, K, for Pt.
Inset shows the Seebeck coefficient of the GeH/Ge(111) film harnessed with Ti electrodes. }
\vspace{-5mm}
\label{fig2}
\end{figure}

Transport properties of the GeH/Ge(111) films were studied using different electrode metals. 
The carrier type was determined from thermoelectric measurements. 
Pt electrodes allowed metallic conduction at $T\!=$20-250\,K (Fig.\,\ref{fig2}).
Residual-resistance ($R_{\rm s}$) ratio, RRR, between 250\,K and 20\,K was $\approx\,$20. 
Hole conduction was confirmed using Seebeck coefficients, $S$ (See inset in Fig.2). 
This marks the first metallic hole conduction reported in GeH/Ge(111), which compares with the semiconducting GeH-EDLT under hole accumulation\cite{36}. 
Notice that metallic transport of electrons, as opposed to holes, had already been observed at ${\rm RRR}\!\!=\!\!1.1$-$1.7$ using Ti electrodes\cite{36}. 

As seen in Fig.\,\ref{fig2}, however, Ti sets the lower limit $T=60$\,K on  $R_{\rm s}$ measurement while 100 K on Hall measurements, which raises an issue. 
Here we show that Pt is better placed than Ti, Ni or Pd, to assess hole conduction by GeH/Ge(111) down to $T=$15 K.
This likely owes to the large work function of Pt (See supporting information). 
In contrast, GeH/Ge(111) became semiconducting at $T\!>\!250$ K. 
Parallel conduction through the Ge substrate and buffer is held responsible, which is supported by Hall measurements. 
On the other hand, we found $S\approx\,$200 $\mu$VK$^{-1}$ for $T$=50-250 K (Fig.\,\ref{fig2}). 
This is consistent with the theoretical value, $\approx$\,200 $\mu$VK$^{-1}$, for a single-layer GeH with sheet carrier density of 2.5$\times10^{12}\, {\rm cm}^{-2}$, and even with the experimental one, $\approx$\,200\,$\mu$VK$^{-1}$, for a p-Ge thin film with sheet carrier density of $\approx$ 1$\times10^{12}\, {\rm cm}^{-2}$ (assuming 1-nm thickness) in the literature \cite{40,41}. 

\begin{figure}[!b]
\begin{center}
\vspace{-5mm}
\includegraphics[width =1.0 \linewidth]{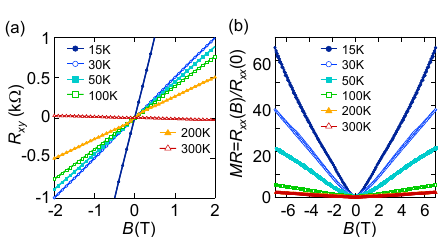}
\end{center}
\vspace{-4mm}
\caption{(a) Hall resistance ($R_{xy}$) and (b) magnetoresistance (MR) as functions of $B$ at various temperatures, 15-300 K.}
\vspace{-5mm}
\label{fig3}
\end{figure}

To determine the carrier density  $n$ and mobility $\mu$, Hall resistance, $R_{xy}(B)$, and magnetoresistance, MR\,$\equiv\!(R_{xx}(B)\!-\!R_{xx}(0))\!/\!R_{xx}(0)$, were measured using Pt electrodes configured in a Hall bar. 
As visible in Fig.\,\ref{fig3}(a), $R_{xy}$ versus $B$ shows positive slopes except at 300 K.  
This along with thermoelectric measurements imply hole conduction. 
The 15-K positive MR reached 6,500 \% at 7 T accompanying minor oscillations (Fig.\,\ref{fig3}(b)). 
Quadratic in $B$ regardless of $T$, MR follows $(\mu B)^2$. 
$\mu$ obtained thereby corresponds to $\mu_{\rm H}$. 
This implies the Lorentz force as the primary cause of MR, which is supported by the anticipated high $\mu_{\rm H}$ due to GeH/Ge(111) AXI. 

Above 200 K, $R_{xy}$ depends nonlinearly on $B$.
This implies two-carrier transport. 
The theoretical fit of $R_{xy}(B)$ allows to distinguish high-mobility holes of GeH/Ge(111) AXI from electrons in the Ge buffer. 
Meanwhile, the estimated $n$ in Ge drops sharply from 10$^{15}$ to 10$^{13}$ cm$^{-2}$ whilst $\mu$ jumps from 10 to 100 cm$^{2}$V$^{-1}$s$^{-1}$ as $T$ varies from 300 to 200\,K (Figure S3). 
This is explicable as due to freeze-out of electrons. 
The Arrhenius plot of $n$, provides the activation energy $E_{\rm a}\!\approx$\,350\,meV. 
Although comparable with $E_{\rm a}$ in non-doped Ge(111), this is an order of magnitude larger than the ionization energy of impurities in Ge, e.g., P or As, $\approx$\! 10\,meV. 
The migrating donor ions from the Ge substrate might have accumulated in the Ge buffer to supply electrons. 
On the other hand, Ge's contribution to conductivity decreases with lowering $T$, accounting for
$<\!10 \%$ of the highest $\mu$ of the AXI channel at 200\,K (Figure S3). 
Hence high-mobility single-carrier transport is most likely to account for the hole conduction in GeH/Ge(111).

\begin{figure}[!b]
\begin{center}
\vspace{-3mm}
\includegraphics[width =1.0 \linewidth]{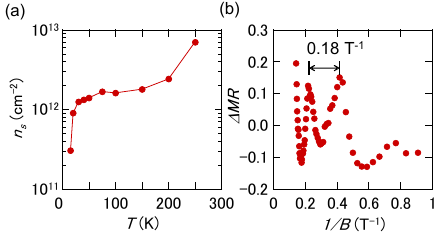}
\end{center}
\vspace{-4mm}
\caption{(a) Temperature dependence of $n_{\rm s}$ obtained by Hall measurements for GeH/Ge(111) thin films with Pt electrodes. 
Hall carrier densities are found using the single-carrier model below 200 K whilst the two-carrier model was used above 200K. 
(b) The oscillatory part of MR ($\Delta$MR) versus $B^{-1}$ at 15 K.  }
\vspace{-3mm}
\label{fig4}
\end{figure}

Sheet carrier density $n_{\rm s}$ was obtained from the single-carrier model, $n_s\!=\!(eR_{\rm H})^{-1}$, 
where $e$ is the elementary charge, and $R_{\rm H}$ the Hall coefficient. 
%$n_{\rm s}$ was then compared with the hole density $h$ available from the MR oscillations at 15 K. 
Figure\,\ref{fig4}(a) shows $n_{\rm s}(T)$ available from Hall resistance whilst (b) shows the oscillatory part of MR, $\Delta$MR, as a function of $B^{-1}$ at 15 K. 
$n_{\rm s}$ using Pt electrodes decreases with decreasing $T$ from $2\!\times\!10^{12}\,{\rm cm}^{-2}$ at 200 K to $3\!\times\!10^{11}\,{\rm cm}^{-2}$ at 15 K. 
Notice that $\Delta$MR equals the as-measured 15-K MR with the second-order polynomial accounting for the background ($\propto\!B^2$) removed.
Oscillations against  $B^{-1}$ are clearly visible (Fig.\,\ref{fig4}(b)). 
This implies the well-resolved Landau levels $E_m\!=\!(m\!+\!\frac{1}{2})\hbar\omega_{\rm c}\ (m\!\ge\!1)$ developing such that $\omega_{\rm c} \tau\!\gg\! 1$  where $\omega_{\rm c}$ is the cyclotron frequency and $\tau$ is the single-particle relaxation time. 
One finds the oscillation period $\Delta(1/B)\!\approx\!\,0.18\,T^{-1}$. 
A spin-degenerate Fermi sphere implies $n_{\rm s}({\rm SdH})\!=\!2e/\{h\Delta(1/B)\}$ where $h$ is the Planck constant. 
From this, we find $n_{\rm s}({\rm SdH})\!\approx\! 3\!\times\!10^{11} {\rm cm}^{-2}$ that aligns with ${n}\!=\!3\times10^{11} {\rm cm}^{-2}$,
which again implies single-band hole conduction along the GeH/Ge(111) AXI.

\begin{figure}[!t]
\begin{center}
\vspace{-5mm}
\includegraphics[width =1.0 \linewidth]{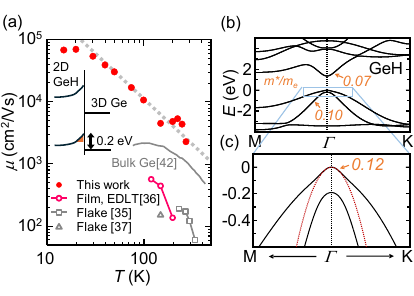}
\end{center}
\vspace{-5mm}
\caption{(a) Hall mobility ($\mu_{\rm H}$) versus temperature of GeH/Ge(111) interface with Pt electrodes (red closed circles).
$\mu_{\rm H}$ of EDLT-GeH thin film (red open circles)\cite{36}, the field-effect mobility of GeH flakes (gray open squares and a triangle) \cite{35,37} and $\mu_{\rm H}$ of p-Ge with $h\!=\!1.2\times10^{17} {\rm cm}^{-3}$(gray curve)\cite{43} are also included for comparative purposes. 
$\mu_{\rm H}$ is estimated by assuming single(two)-carrier model below(above) 200 K. 
The gray dotted line with a negative slope -3/2 implies phonon scattering. 
(b) Band structure of multi-layered GeH with spin-orbit coupling.
(c) The magnified view of the valence band maximum (VBM). 
The hole effective mass $m_{\rm h}^*$ and electron one $m_{\rm e}^*$ in units of the electron mass $m_{\rm e}$ are shown in orange. 
The red dashed curve is a parabolic fit with $m_{\rm h}^*=0.12m_{\rm e}$.}
\vspace{-5mm}
\label{fig5}
\end{figure}

In Fig.\,\ref{fig5}(a), negative ${\rm d}\mu/{\rm d}T$ occurs between $2\!\times10^{3}\, {\rm cm}^{2}{\rm V}^{-1}{\rm s}^{-1}$ at 200 K and record high 6.7$\times$10$^4$ ${\rm cm}^{2}{\rm V}^{-1}{\rm s}^{-1}$ at 15 K. 
This thermal roll-off follows $\mu\!\propto\!T^{-3/2}$ for $T\!>\!30$\,K, which implies the relevance of acoustic phonon scattering. 
Meanwhile, a plateau develops on the low-$T$ side down to 15\,K.
These remind us of high-mobility modulation-doped 2DEG/2DHG heterointerfaces/structures\cite{1,42} as a result of suppressed ionized impurity scattering.
This implies that holes in the GeH/Ge(111) AXI are barely susceptible to ionized acceptor scattering. 
This contrasts with the bulk where positive ${\rm d}\mu/{\rm d}T\!>\!0$ brings a peak of $\mu(T)$.
The bleached screening of Coulomb potential due to deactivated thermal carriers is responsible.
$\mu(T)$ of bulk p-Ge  ($n\!\approx$10$^{17}$cm$^{-3}$) is reproduced in Fig.\,\ref{fig5} by the gray curve\cite{43}. 

Next we discuss the enhanced $\mu$.
To date, $\mu\!=\!150$\,cm$^{2}$V$^{-1}$s$^{-1}$ at 150 K for GeH flakes\cite{35} and $\mu\!=\!570$\,${\rm cm}^{2}{\rm V}^{-1}{\rm s}^{-1}$ at 120 K for GeH/Ge(111) thin-film EDLTs are known\cite{36}. 
We explore conceivable factors that allow high mobilities.
First, the quality of in-plane crystallinity.
The epitaxial GeH on Ge(111) outperforms GeH flakes on SiO$_2$/Si prepared by exfoliation or evaporation drying of a GeH-laden solution\cite{35,37}. 
Second, the relevance of a special conducting channel.
This is probable as GeH/Ge(111) films exhibit hole conduction without FET-geometry, unlike the previous insulating ones.
%Other differences are attributable to the growth conditions. 
Now the question is what and where such a putative channel is:
A hypothesized 2D conductor along the surface or GeH/Ge(111) AXI or a 3D one through GeH are conceivable. 
In the same context, we discuss doping. 
Importantly, metallic GeH disproves 3D conduction as discussed below. 
The insulator-to-metal transition occurs according to the Mott criterion at the critical density $n_{\rm c}$. 
Here $n_{\rm c}^{1/3}a_{\rm B}^*$\!=0.26 with $a_{\rm B}^*$ being the effective Bohr radius \cite{44}. 
For $n\!\approx \!3\times10^{16} {\rm cm}^{-3}$ at 15 K and 100-nm thickness, the relative dielectric constant of GeH $\approx$3.7 \cite{45} and the hole effective mass 0.07 $m_{\rm e}$ with $m_{\rm e}$ being the electron mass, one finds $n_{\rm c}\!=\!8.0\times 10^{17}  {\rm cm}^{-3}$. 
The observed $n\!\approx \!3\times10^{16} {\rm cm}^{-3}$ is thus too low to explain in the context of 3D metal i.e., degenerate semiconductors. 

Therefore the hole channel, if exists, must be localized at the surface or interface. 
Previously, we realized 2DHG on GeH underby FET-action with
$R_{\rm s}$(200 K)$\approx$ 20 k$\Omega$ upon hole accumulation\cite{3}. 
On the contrary, hole accumulation here induces a metallic behavior with $R_{\rm s}$(200 K)$\approx$1.7 k$\Omega$.
In light of the comparable crystallinity of GeH, where holes accumulate, plays a critical part. 
Rather than exposed at the surface, the conducting channel is likely to be  located at the GeH/Ge(111) AXI.
This is because epitaxy ensures well-ordered crystalline interfaces with good electrical contacts among crystal grains compared to the surface. 
To substantiate such a conjecture, the top part of GeH was dry-etched in steps after transport measurements. 
Resistivity and Hall resistance remained unchanged down to 20 K as expected.
This implies that the conducting channel is localized near GeH/Ge(111) interface, not elsewhere (Figure S5). 
It is most likely that a modulation-doped heterostructure is formed between 2D GeH and 3D Ge(111), i.e., at the GeH/Ge(111) AXI  (Fig.\,\ref{fig5}(a)). 

To gain insight into this, we computed their potential alignment and the band dispersion of the GeH.
The work function of the GeH was calculated to be 4.45 eV. This is lower than the calculated value of 4.65 eV for a GeH (111) surface with the same pseudopotential and the experimental value of 4.8 eV.\cite{47} This indicates that the GeH is responsible, and the potential alignment is type-II, as illustrated in the inset of Fig. 5(a). This implies that holes are transferred from Ge to GeH and the 2DHG is isolated from ionized acceptors in the Ge buffer.
In Ge, 2D confinement under compressive stress results in a small hole effective mass, $m_{\rm h}^*$. 
In contrast, the inherent 2D nature of GeH permits a small effective mass. 
It is noticed that a direct gap opens ($E_{g}\!\approx\!$1.4 eV) (Fig.\,\ref{fig5}(b) and Figure S6). 
In view of this, near-edge bands are hereafter labelled with respect to the valence band maximum (VBM). 
The first band is made of Ge p-orbitals with total angular momentum $j\!=\!3/2$ with its projection along the quantization axis $j_z\!=\!\pm 3/2$, i.e., $\ket{j, jz}\!=\!\ket{3/2, \pm 3/2}$. 
The second band, 0.2-eV below VBM, consists of $\ket{1/2, \pm 1/2}$ and $\ket{3/2, \pm 1/2}$ of the Ge p-orbitals. 
The third flat band 3-eV below VBM is a hybrid of the H s-orbital and $\ket{3/2, \pm 1/2}$ and $\ket{1/2, \pm 1/2}$ of the Ge p-orbital, i.e., the out-of-plane basis of the GeH sheet. 
The overall dispersion is consistent with the previous studies \cite{24,25,46}. 
A large curvature of the first band is noticed within 0.1 eV below VBM  (Fig.\,\ref{fig5}(c) and Figure S6). 
$m_{\rm h}^*\!=\!0.14m_{\rm e}$ is found near the $\Gamma$-point shown in Fig.\,\ref{fig5}(c) after curve fitting. 
Meanwhile, experiments predict $n\!=\!10^{19}$\,cm$^{-3}$ for the GeH/Ge(111) AXI ($\approx$\,1\,nm). 
Referring to Figure S6(c), such a hole density implies the exclusive occupation of the first band. 
The calculation further implies single-band hole conduction along the isotropic 2D quasi-Fermi surface. 

From the foregoing discussion, it is judicious to conclude that the high hole mobilities observed for the GeH/Ge(111) allotropic cross-D heterointerface are consistent with not only the small scattering rate but also the small effective mass in the GeH conducting layer in a type-II band alignment.

%\section*{\large{Conclusion}} 　
%In summary, the allotropic cross-D heterointerface between the 2D hydrogen-terminated layered germanium and Ge(111) prepared by topotactic reaction provided higher-than-expected hole mobility and well-resolved Landau levels at 15 K. 
%The Hall mobility reached 67,000 ${\rm cm}^{2}{\rm V}^{-1}{\rm s}^{-1}$ with its temperature dependence implying a strong suppression of ionized acceptor scattering. 
%A high quality of crystallinity at the bottom interface of the GeH film should explain all. 
%Spontaneous modulation-doping of holes was attributed to the work function difference between GeH and Ge. 
%The allotropic cross-D heterinterface between 2D van der Waals and traditional 3D bulk
%opens the avenue towards facile 2DHG creation without the need for elaborate heteroepitaxy and precision modulation doping. 
%Further attempts at lower temperatures is likely to allow quantum Hall effect in this yet unexplored class of semiconductor interface.

%Acknowledgments
This work was in part supported by JSPS KAKENHI JP21H01038, JP22K14001, JP22K04869, JP20H02635, JP23H01865, JP23K26558, JP25K01617 and 25K17340.This work was also partly supported by the Iketani Science and Technology Foundation, and the Toyota Physical and Chemical Research Institute.

\end{document}